\documentclass[aps,prb,twocolumn,amsmath,amssymb]{revtex4-1}
\usepackage{times}
\usepackage{graphicx}
\usepackage[caption=false]{subfig}
\usepackage{amsfonts}
\usepackage{amsmath, amsthm, amssymb}
\usepackage{textcomp}
\usepackage{color}
\usepackage{bm}
\usepackage{float}
\usepackage{soul}
\usepackage{siunitx}
\definecolor{DarkRed}{rgb}{0.65,0,0}%
\definecolor{DarkBlue}{rgb}{0,0,0.65}

\newcommand{\Real}{{\mathrm{Re}}}   
\newcommand{\ve}[1]{\boldsymbol{#1}}
\DeclareMathOperator{\diag}{diag} 

\newcommand{\vech}{\ve{h}} 

\def\i{\mathrm{i}}

\begin{document}

\title{Supercurrent Vortex Pinball via a Triplet Cooper Pair Inverse Edelstein Effect}

\author{Morten Amundsen and Jacob Linder}

\affiliation{QuSpin Center of Excellence, Department of Physics,  \\Norwegian University of Science and Technology, N-7491 Trondheim, Norway}

\begin{abstract}
We consider the Josephson effect through a thin spin-orbit coupled layer in the presence of an exchange field $\vech$, and discover a set of supercurrent vortices appearing in the system which can be controllably moved around in the system by varying either the direction of $\vech$, the strength $|\vech|$, the spin-orbit coupling magnitude $\alpha$ via a gate voltage, or the phase difference. We refer to this phenomenon as a supercurrent vortex pinball effect and show that its origin is the spin polarization of the triplet Cooper pairs induced in the system. The supercurrent vortices are shown to arise from what resembles a Cooper pair-induced inverse Edelstein effect.  
\end{abstract}

\date{\today}

\maketitle
\section{Introduction}
The proximity effect endows otherwise non-superconducting materials with superconducting properties, and thus offers an interesting arena to explore how quantum coherence is manifested in unusual physical environments. One of the most actively pursued setups includes so-called Josephson junctions where a material is able to sustain a supercurrent via proximity to two host superconductors. 
In the presence of a magnetic exchange field $\vech$, the Cooper pairs that leak from the superconductors oscillate between the conventional singlet spin state and the opposite-spin triplet state \cite{buzdin_rmp_05}. If the exchange field is spatially varying, or the system contains spin-orbit coupling, the triplet Cooper pairs may rotate between different triplet states where the electron spins are parallel, making tunable superconducting spin transport possible~\cite{bergeret_prl_01,linder_nphys_15, eschrig_rpp_15}. In light of this discovery, the effect of spin-orbit coupling on proximitized materials has recently been investigated in several works~\cite{bergeret_prl_13, jacobsen_prb_2015, jacobsen_prb_2015b, konschelle_prb_15, arjoranta_prb_16, alidoust_njp_15, pershoguba_prl_15, konschelle_prb_2016, costa_arxiv_16, bobkova_arxiv_16, takashima_prb_16, jacobsen_scirep_15, yu_prb_16, tkachov_prl_17, pientka_prx_17, bjornson_prb_15, malshukov_prb_16}.

 In this paper, we consider a spin-orbit coupled superconducting hybrid which is found to display novel, inherently two-dimensional, physical phenomena which are hidden in effective one-dimensional models. Specifically, we examine a thin film of a non-superconducting material with spin-orbit coupling sandwiched between two superconductors and discover a set of supercurrent vortices appearing in the system which can be controllably moved around by varying either the direction of $\vech$, the strength $|\vech|$, the spin-orbit coupling magnitude $\alpha$ via a gate voltage or the phase difference. We refer to this phenomenon as a supercurrent vortex pinball effect and show that its origin is the spin polarization of the triplet Cooper pairs induced in the system. The supercurrent vortices arise from what is reminiscent of a Cooper pair-induced inverse Edelstein effect.

\begin{figure}[b!]
\includegraphics[width=0.5\columnwidth, angle=-90]{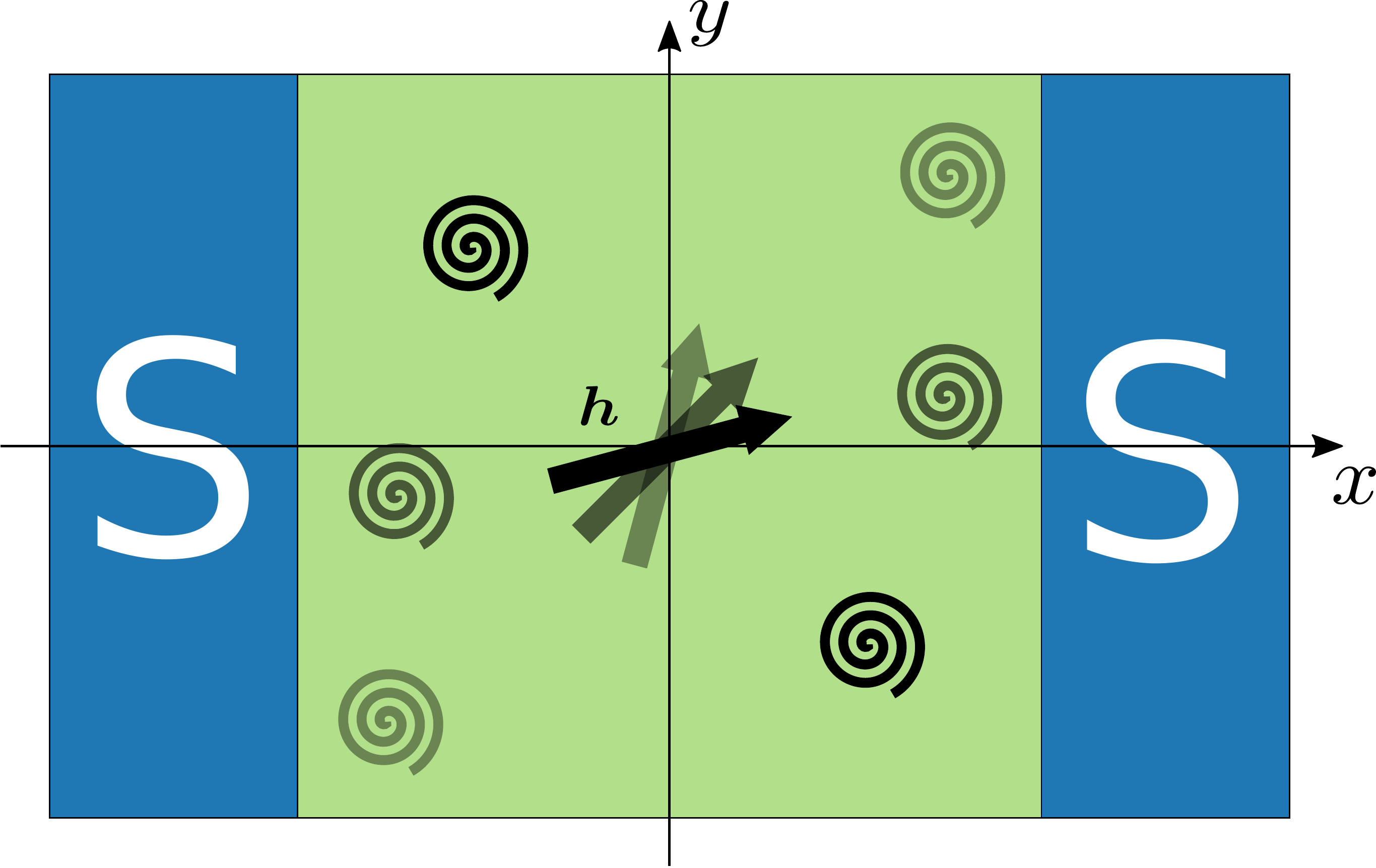}
\caption{A top view sketch of the 2D Josephson junction, which is oriented in the $xy$-plane. The green region reprsents the normal metal, which is quadratic with lengths $L = 2\xi$. The exchange field $\bm{h}$ is applied in the plane of the junction. The presence of spin-orbit coupling creates vortices in the current density.}
\label{fig:sketch}
\end{figure}

\section{Theory and model}
A top view sketch of the geometry is given in Fig.~\ref{fig:sketch}. 
The film can for instance be a two-dimensional electron gas (GaAs), which has the advantage of a readily tunable Rashba spin-orbit coupling strength $\alpha$ and a high $g$-factor providing a strong Zeeman effect. The central region is quadratic with lengths $L = 2\xi$ where $\xi$ is the superconducting coherence length and considered to be in the diffusive regime of transport. A magnetic field is applied in the $xy$-plane, in a direction denoted by an angle $\theta$ relative to the transverse direction ($y$-axis), creating an exchange field $\bm{h}$ through the Zeeman effect. We assume that the film is sufficiently thin [$\mathcal{O}$(nm)] for the orbital effect to be completely negligible.  In the diffusive transport regime, the Usadel equation~\cite{usadel} is valid:
\begin{align}\label{eq:usadel}
D\bar{\nabla}\hat{g}\bar{\nabla}\hat{g} + \i\left[\varepsilon\hat{\rho}_3 + \hat{\bm{\sigma}}\cdot\bm{h} , \hat{g}\right] = 0
\end{align}
where $D$ is the diffusion constant, $\varepsilon$ is the quasiparticle energy, $\hat{\rho}_3 = \diag(+1,+1,-1,-1)$, and $\hat{\bm{\sigma}} = \diag(\bm{\sigma} , \bm{\sigma}^*)$ where $\bm{\sigma}$ is a vector of Pauli matrices. Furthermore, $\hat{g} = \hat{g}(x,y,\varepsilon)$ is the retarded, quasiclassical $4\times4$ Green function matrix, defined as 
\begin{align}
\hat{g} = \begin{pmatrix}
g & f \\
-\tilde{f} & -\tilde{g} \\
\end{pmatrix}
\end{align}
where the $\tilde{\ldots}$ operation means complex conjugation and $\varepsilon \to (-\varepsilon)$. The $ 2\times 2 $-matrix $g$ is the conventional Green function, which includes the spin degree of freedom, whereas the anomalous Green function $f$ takes into account the presence of superconducting correlations. Spin-orbit coupling is introduced via the covariant derivative ${\bar{\nabla}\hat{g} = \nabla\hat{g} - \i\left[\hat{\bm{A}}, \hat{g}\right]}$, with $\hat{\bm{A}} = \diag(\bm{A}, -\bm{A}^*)$. Here, we consider Rashba spin-orbit coupling, as generated by a symmetry breaking in the thickness direction ($z$-axis), for which one gets 
\begin{align}
\bm{A} = -\alpha\left(\sigma_y\bm{e_x} - \sigma_x\bm{e_y}\right)
\end{align}
where $\alpha$ is the strength of the spin-orbit coupling, and $\bm{e_j}$ is a unit vector in direction $j$. We consider low-transparancy interfaces, which are described by the Kupriyanov-Lukichev boundary conditions~\cite{KL} 
\begin{align}
\bm{e}_N \cdot \bar{\nabla}\hat{g} = \frac{1}{\zeta L}\left[\hat{g}_R, \hat{g}_L\right]
\end{align}
 The phenomenological constant $\zeta$ is a measure of the interface resistance (we use $\zeta=3$ in what follows, corresponding to a realistic low-transparency interface), and the indices $\{L, R\}$ refer to Green functions on the left and right side of the interface, respectively. Inelastic scattering is accounted for by letting $\varepsilon \to \varepsilon + \i\delta$ where $\delta/\Delta=0.01$. The superconductors are assumed to be large enough to be approximated as bulk, and therefore appear only in the boundary conditions at $x = \pm L/2$. For the transversal interfaces to vacuum, the boundary conditions reduce to $\bar{\nabla}\hat{g} = 0$. 

To demonstrate the inverse Edelstein effect originating from the triplet Cooper pairs, we first consider the charge current density which in equilibrium is defined as:
\begin{align}\label{eq:current}
\bm{J} = J_0 \int d\varepsilon\;\Real\text{Tr}\left\{ \hat{\rho}_3\hat{g}\bar{\nabla}\hat{g}\right\} \tanh \frac{\beta \varepsilon}{2}
\end{align}
with $J_0 =N_0eD/2$, where the constant $N_0$ is the density of states at the Fermi level and $\beta = 1/k_B T$. The temperature $T$ is constant and equal to {1 \textperthousand} of the critical temperature of the superconductors. By inserting the above expression for $\hat{g}$ into Eq.~\ref{eq:current}, it is seen that the current density only depends on the anomalous Green function $f$. Furthermore, $f$ can be split into a contribution from the singlet component $f_s$ and the triplet component $\bm{f}_t = \left(f_x, f_y, f_z\right)$ by inserting the parameterization 
\begin{align}
f = \left(f_s I + \bm{f}_t\cdot \bm{\sigma}\right) \i \sigma_y
\end{align}
where $I$ is the $2\times2$ identity matrix.  The current density can then be written as
\begin{align}\label{eq:currentsplit}
\bm{J} = \bm{J}_{I} + \bm{J}_{tt}
\end{align}
with $\bm{J}_I = \bm{J}_s - \bm{J}_x - \bm{J}_y - \bm{J}_z$ being the conventional current contribution from the individual singlet and triplet components, which in turn are given as

\begin{align}
\bm{J}_k = 4J_0 \int_0^{\infty} d\varepsilon\;\Real\left\{ \tilde{f}_k \nabla f_k - f_k\nabla\tilde{f}_k\right\}\tanh \frac{\beta \varepsilon}{2} \label{eq:Jk}
\end{align}
for $k\in\{s, x, y, z\}$. The current $\bm{J}_I$ is present also in the absence of spin-orbit coupling. With spin-orbit coupling, however, one gets \textit{an additional contribution} from $\bm{J}_{tt}$, which we find to be:
\begin{align}
\bm{J}_{tt} = 16J_0 \alpha\bm{e}_z\times \int_0^{\infty} d\varepsilon\; \Real \left\{ \tilde{\bm{f}}_t\times\bm{f}_t\right\}\tanh \frac{\beta \varepsilon}{2} \label{eq:Jtt}
\end{align}
It is observed that while $\bm{J}_I$ is a linear combination of currents from each of the four components of $f$, $\bm{J}_{tt}$ is generated by interference between the triplet components. Importantly, the cross product in Eq.~\ref{eq:Jtt} determines the spin polarization direction of  a general triplet Cooper pair state, as is well-known from the $d$-vector formalism used in early works on liquid $^3$He \cite{leggett_rmp_75, maeno_rmp_04}. In other words we find that, due to spin-orbit coupling, the \textit{existence of a finite triplet Cooper pair spin expectation value directly produces a charge current}, which we interpret as a triplet Cooper pair induced inverse Edelstein effect. This is a key result in this paper. We note that another type of Edelstein effect in superconducting hybrid structures has recently been reported in Ref.~\cite{konschelle_prb_15}, where a spontaneous supercurrent induced by magnetization in a Josephson junction with spin-orbit coupling was found. Here, we have presented a different inverse Edelstein effect, in that the spin density responsible for the current is solely generated by the triplet Cooper pairs. This is in contrast to the induced magnetization which requires a non-zero singlet contribution. Recently, the nonequilibrium Edelstein effect and magnetoelectric Andreev transport was discussed in the context of helical metals \cite{tkachov_prl_17}.

\section{Results and discussion}
For the system shown in Fig.~\ref{fig:sketch}, we solve the full Usadel equation, given in Eq.~\ref{eq:usadel}, by using the finite element method, as thoroughly explained in Ref.~\cite{amundsen_scirep_16}. In the results presented herein we apply a phase difference between the superconductors of $\phi = \frac{\pi}{2}$ unless otherwise stated. The first thing to note is that the current density has a non-trivial transversal distribution, and even changes sign in certain areas. In the regions where a sign change occurs, supercurrent vortices are generated, i.e. positions around which the current density circulates. Vortices in the current density have been reported in proximitized materials in the presence of an external magnetic flux~\cite{cuevas_prl_07, bergeret_jlt_08, alidoust_prl_12, amundsen_scirep_16}, whereas no such flux is required in the present work. These flux-induced vortices are associated with a suppression of superconducting correlations at precisely the location of the vortices in addition to a phase-winding of the superconducting phase, analogously to Abrikosov vortices. In the results presented here, we do not find any such suppression, and so the spin-orbit induced supercurrent vortices are therefore of a different nature.   
\begin{figure}[t!]
\subfloat[]{%
\includegraphics[width=0.51\linewidth]{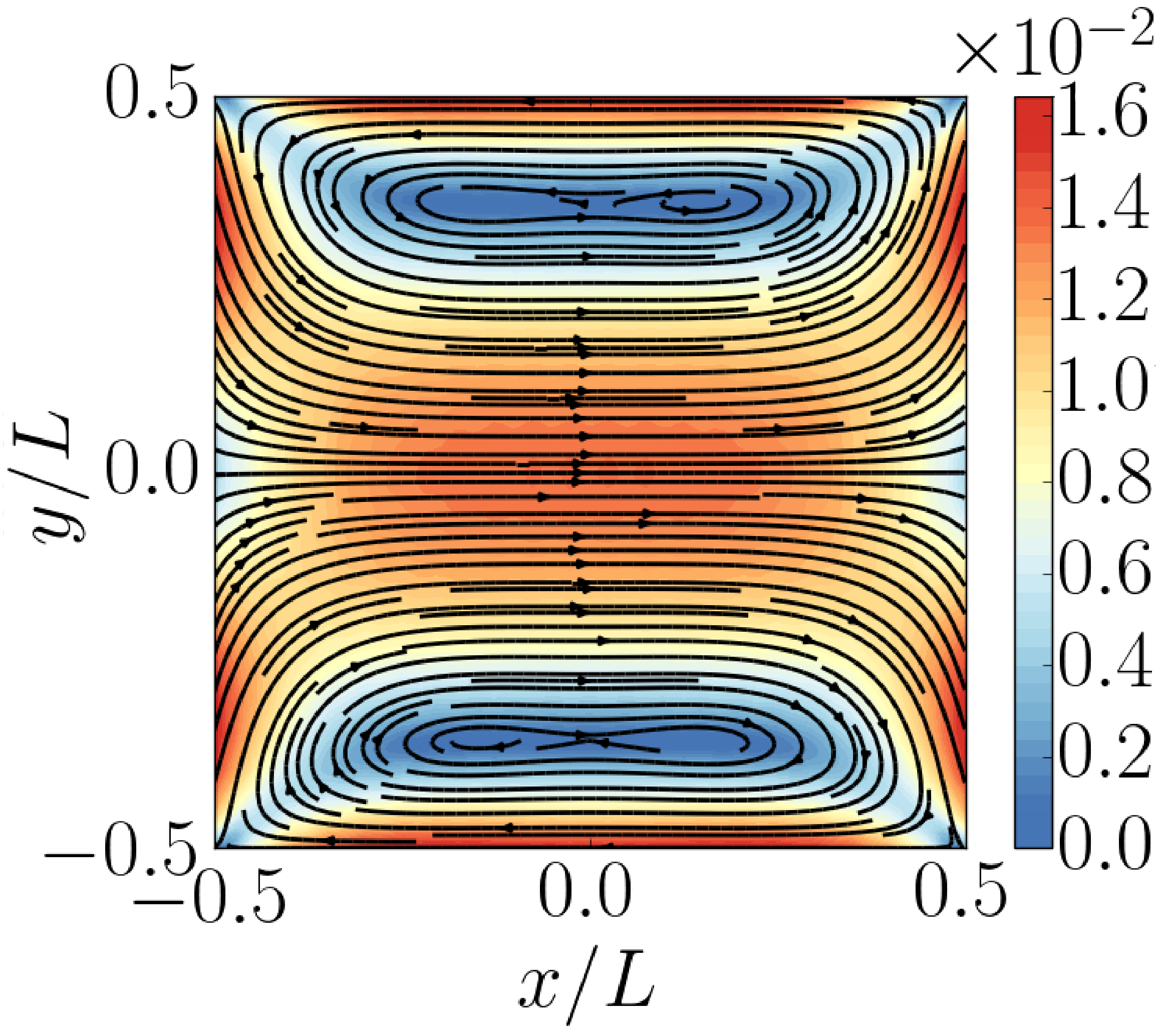}
\label{fig:hxa}
}
\subfloat[]{%
\includegraphics[width=0.49\linewidth]{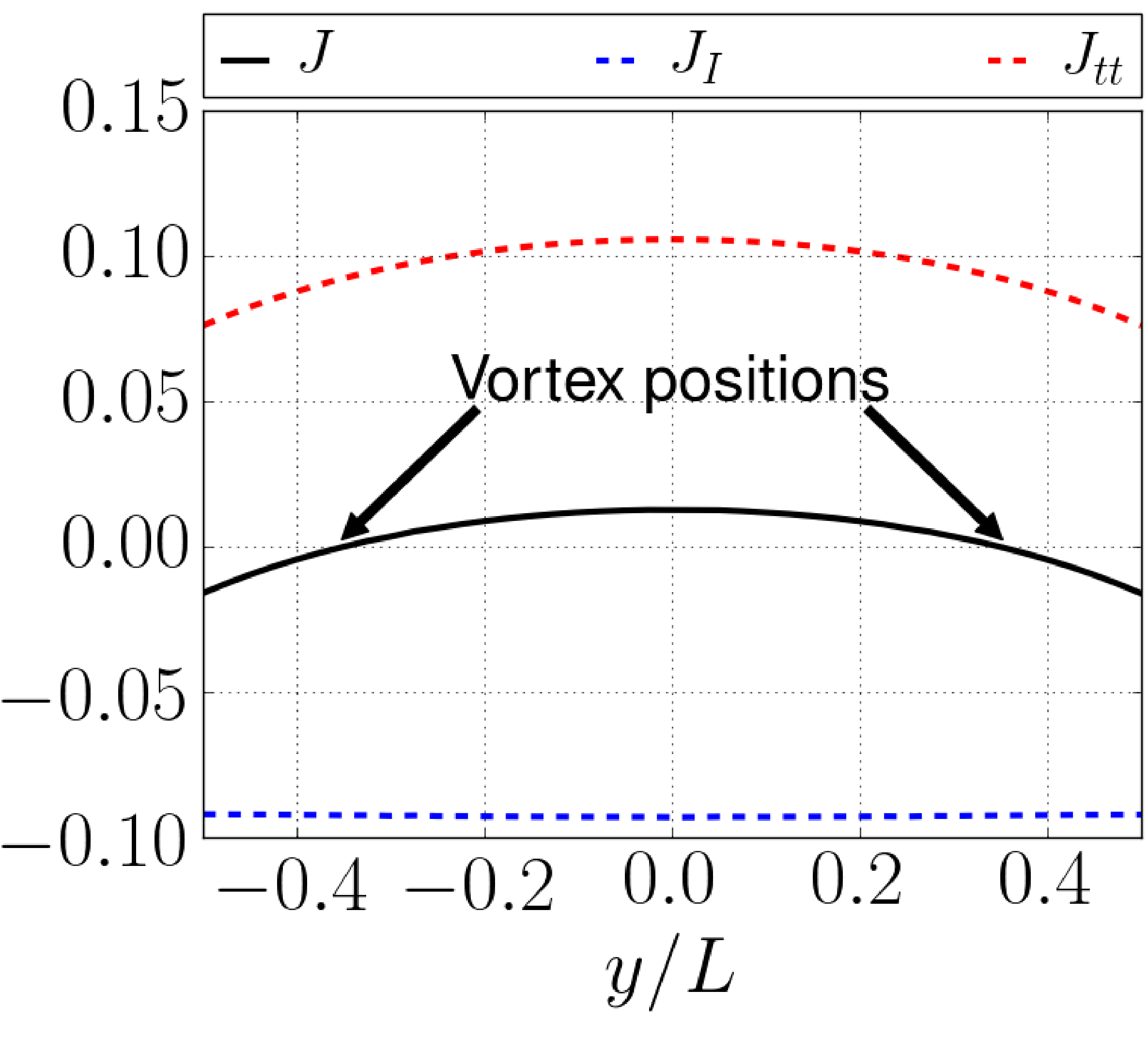}
\label{fig:hxb}
}
\caption{The current density distribution for an exchange field $h/\Delta = 1.0$ in the $x$-direction. (a) Streamline plot showing vortices. (b) Transversal distribution of the $x$-component of the different currents in the middle of the junction ($x = 0$). All current densities are scaled by $J_0\Delta/L$.}
\label{fig:hx}
\end{figure}

The effect is clearly seen in Fig.~\ref{fig:hxa}, in which the exchange field is pointing in the $x$-direction. Here, the current density flows in the positive $x$-direction near the center of the junction, and in the opposite direction by the edges. This creates a circulation around two oblong vortices at approximately $y = \pm 0.4L$  within which the current density is suppressed. Such a current distribution may be measured experimentally using magneto-optic imaging~\cite{johansen_prb_96}. 

Fig.~\ref{fig:hxb} shows the transverse distribution of the current flowing across the junction. Specifically, it shows the different contributions in Eq.~\ref{eq:currentsplit} to the total current. It is seen that the transverse distribution of the individual components $J_k$---the components not explicitly dependent on $\alpha$---almost perfectly cancel, rendering their sum $J_I$ constant. It is therefore clear that the main contribution to the transverse distribution stems from $J_{tt}$, the term responsible for the inverse Edelstein effect induced by the triplet Cooper pair spins.

\begin{figure}[t!]
\centering
\includegraphics[width=\linewidth]{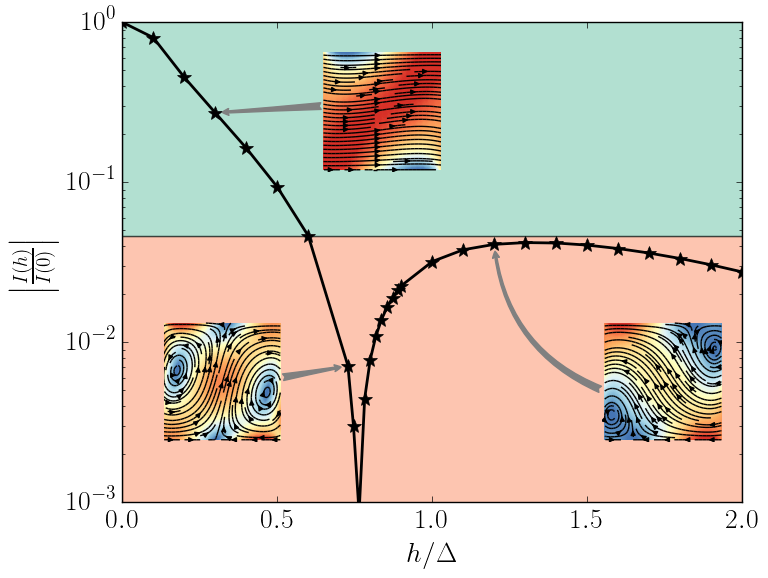}
\caption{The total current $I$ passing between the superconductors (found by integrating the current density $\bm{J}$ over the cross section) as a function of the exchange field strength $h=|\vech|$, applied at an angle $\theta = 45^{\circ}$ relative to the $y$-axis. The normalized strength of the spin-orbit coupling is $\alpha L = 1$. Pink regions indicate presence of vortices in the current density. The insets show the current density distribution for selected points.}
\label{fig:current}
\end{figure}

A key point is that the \textit{existence} of the vortices are found to depend on the strength $h=|\vech|$ of the exchange field. When examining the current passing between the superconductors for increasing exchange field, a decaying oscillatory pattern is found where, for certain values of $h$, the current flows in the opposite direction. This is known as a $\pi$-junction \cite{ryazanov_prl_01}, owing to the fact that the ground state of the Josephson junction has a phase difference of $\pi$ rather than zero~\cite{buzdin_rmp_05}. We find that the vortices are harbingers of a 0-$\pi$ transition, being present only when the total current is significantly reduced. The reason for this is that the transversal distribution of the current density is much less influenced by the strength of the exchange field than the current itself. For increasing $h$, the curvature of $J_{tt}$ in Fig.~\ref{fig:hxb} is more or less retained, while the total current---that is, the average of the current density---is reduced upon approaching a 0-$\pi$ transition. Close enough to the transition, the minimas of the current density will cross zero and become negative. Evidently, a current density redistribution is less energetically favorable than vortex generation.

It turns out that for an in-plane exchange field, the current is reduced rapidly enough with increasing $h$ for vortices to remain present once they first appear. This can be seen in Fig.~\ref{fig:current}, which shows the total current $I$, found by integrating the current density $\bm{J}$ over the cross section. The exchange field is applied at an angle $\theta=45^{\circ}$ with respect to the $y$-axis. In the pink region there are vortices in the current density, and it is observed that the modulation of the curve hinders a reentry into the green region, where vortices are not present. For an out-of-plane exchange field this is not necessarily the case, and away from the transition points, the current may become large enough for the vortices to disappear.

Interestingly, the \textit{location} of the vortices also changes as the exchange field is increased. The insets of Fig.~\ref{fig:current} show the current density distribution for selected points along the current curve. The vortices first appear at the vacuum edges, on opposite sides of the junction, near the superconducting interfaces. As $h$ is increased, they translate vertically and pass the $x$-axis at \textit{precisely} the 0-$\pi$ transition. At this stage, the current density distribution is symmetric about both the $x$ and the $y$ axis, with no net current passing between the superconductors. Further increase of $h$ causes further translation of the vortices. However, since they must cross the $x$-axis every 0-$\pi$ transition due to the symmetry requirements, a turning point must be reached, and the motion of the vortices may best be described as resembling a damped harmonic oscillator.

\begin{figure}
\subfloat[$\theta=30^{\circ}$]{%
  \includegraphics[width=0.5\columnwidth]{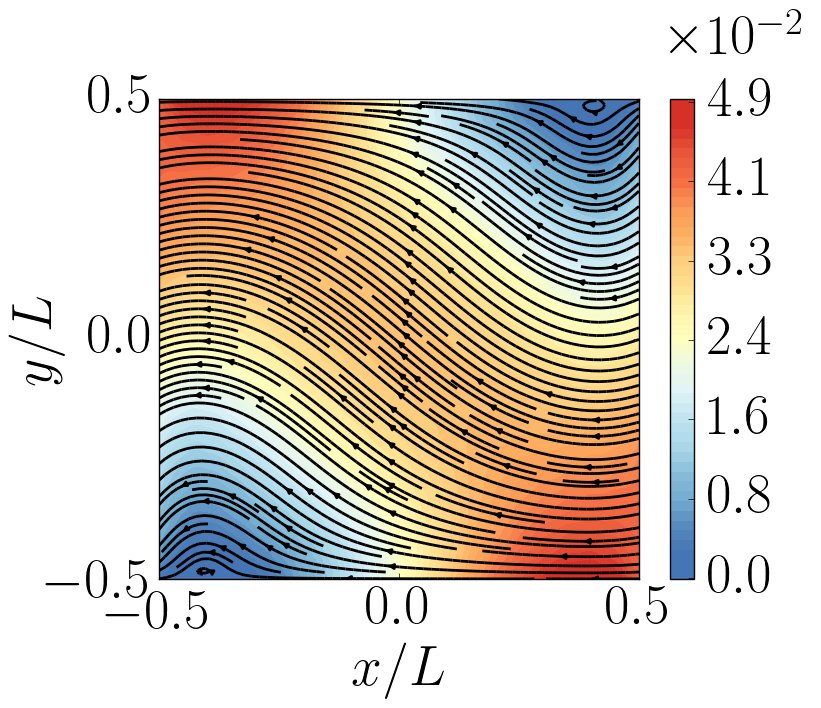}
\label{fig:hxy1}
}
\subfloat[$\theta=60^{\circ}$]{%
  \includegraphics[width=0.5\columnwidth]{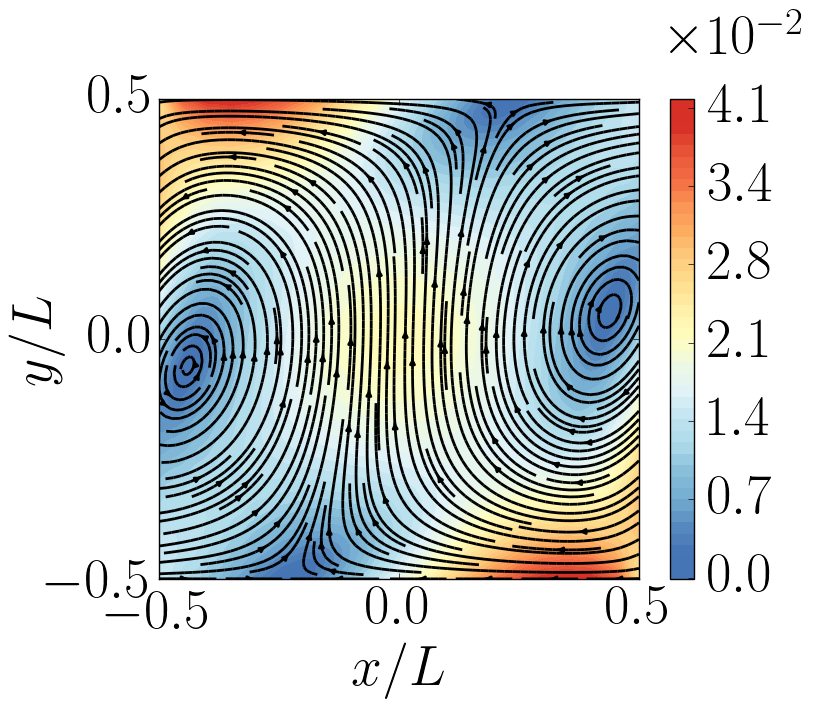}
\label{fig:hxy2}
}

\subfloat[$\alpha L = 1.2$]{%
  \includegraphics[width=0.5\columnwidth]{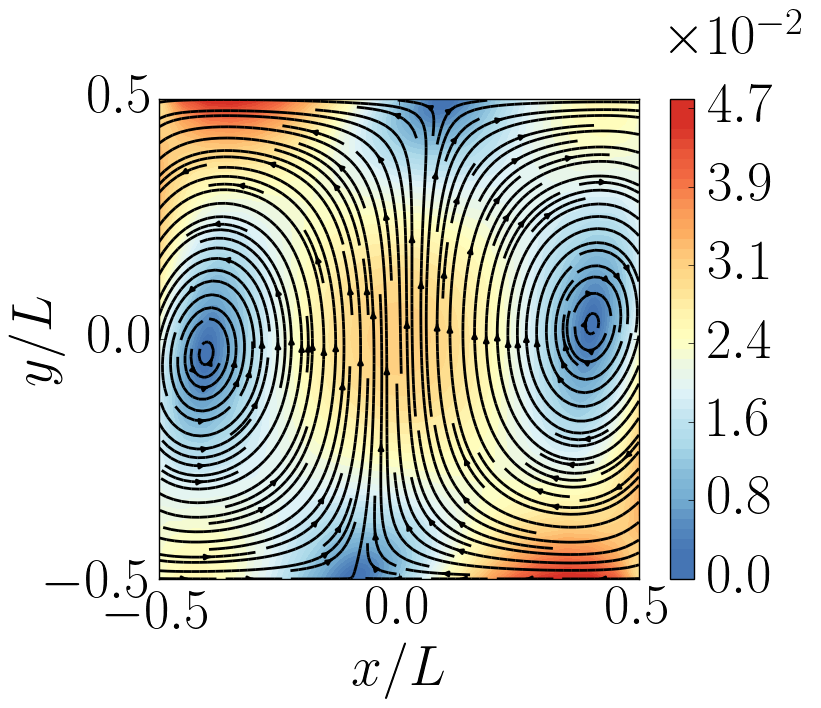}
\label{fig:hxy3}
}
\subfloat[$\alpha L = 1.6$]{%
\includegraphics[width=0.5\columnwidth]{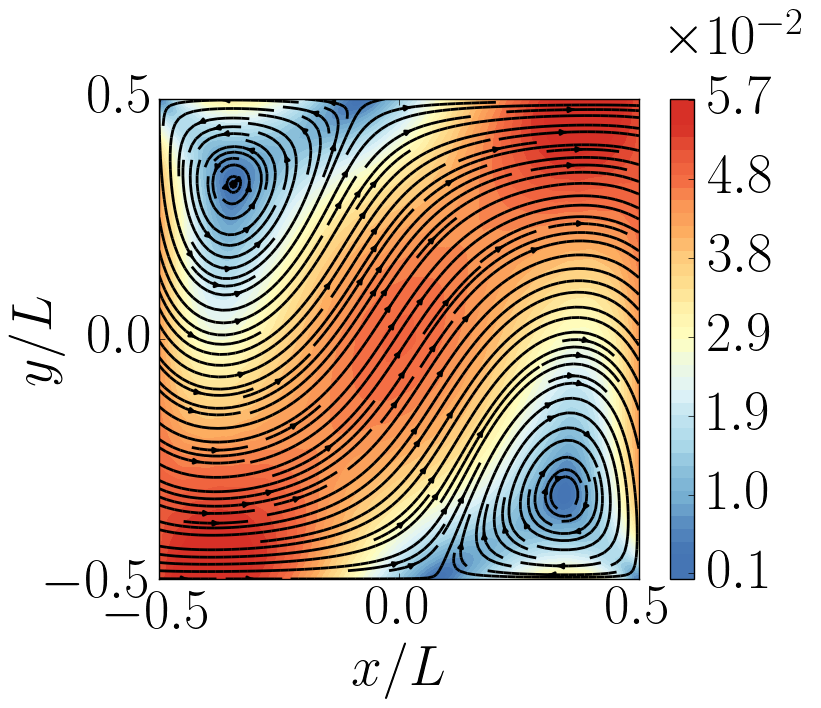}
\label{fig:hxy4}
}

\subfloat[$\phi = \frac{\pi}{5}$]{%
  \includegraphics[width=0.5\columnwidth]{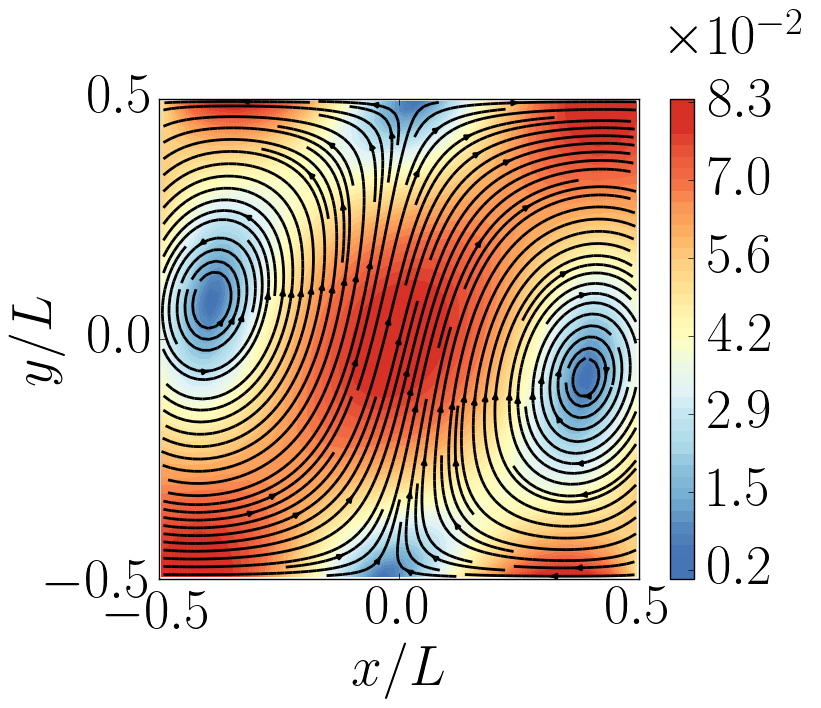}
\label{fig:phi1}
}
\subfloat[$\phi = \frac{4\pi}{5}$]{%
\includegraphics[width=0.5\columnwidth]{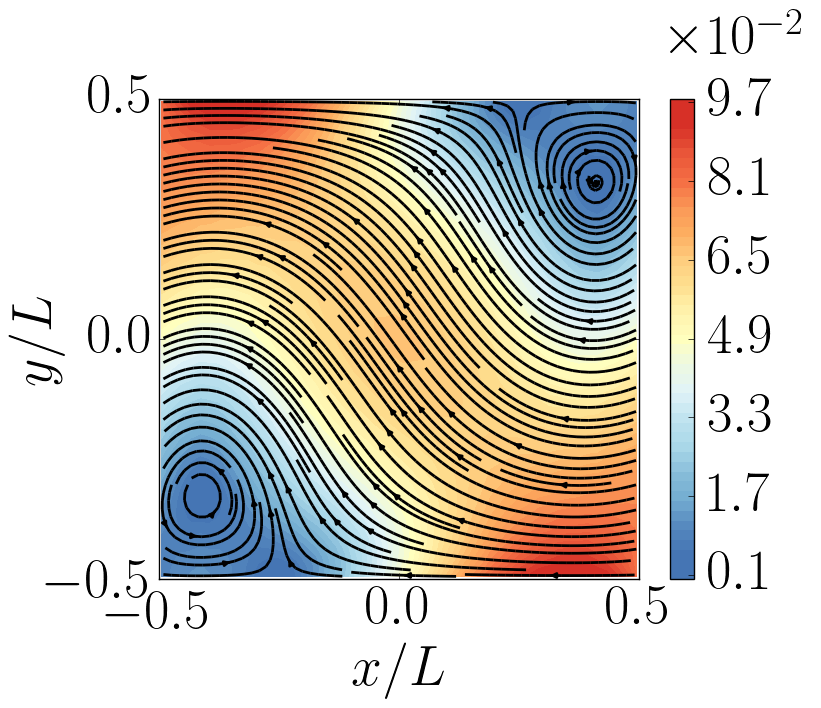}
\label{fig:phi2}
}
\caption{Streamlines of the current density for: (a)-(b) Different exchange field directions $\theta$ with spin-orbit coupling strength $\alpha L = 1.0$ and phase difference $\phi = \frac{\pi}{2}$. (c)-(d) Different $\alpha L$ with $\theta = 45^{\circ}$ and $\phi = \frac{\pi}{2}$. (e)-(f) Different $\phi$ with $\alpha L = 1.0$ and $\theta = 45^{\circ}$. The strength of the exchange field is $h/\Delta = 1.0$,  which is a realistic magnitude of order $\simeq$ meV accessible via an external magnetic field. All current densities are scaled by $J_0\Delta/L$.}
\label{fig:hxystream}
\end{figure}

In the study of a 1D Josephson junction with spin-orbit coupling, it was recently discovered that the critical current varies greatly with the direction of an applied exchange field, even creating 0-$\pi$ transitions~\cite{arjoranta_prb_16, costa_arxiv_16}. It is therefore reasonable to presume that the current density distribution becomes non-trivial.
 We find that this is indeed the case. With an exchange field of strength $h/\Delta = 1.0$ pointing in the transversal direction ($\theta = 0^{\circ}$), no vortices are found. Rotating $\bm{h}$ increases the transversal variation of the current density, and at around $\theta = 30^{\circ}$ vortices appear, as shown in Fig.~\ref{fig:hxy1}. Further rotation translates the vortices vertically towards the $x$-axis, as seen in Fig.~\ref{fig:hxy2} for $\theta = 60^{\circ}$, before translating towards $y$-axis, and ending up like Fig.~\ref{fig:hxa} for $\theta=90^{\circ}$.

We also determine how the strength of the spin-orbit coupling affects the vortices. Tuning of the Rashba parameter $\alpha$ can be achieved experimentally in a 2DEG by means of a gate voltage~\cite{nitta_prl_97, engels_prb_97, grundler_prl_00}. 
For small spin-orbit coupling strength, so too is the contribution from $J_{tt}$, as can be seen from the explicit dependence on $\alpha$ in Eq.~\ref{eq:Jtt}. With increasing $\alpha$, the inverse Cooper pair Edelstein effect predicted here increases both in terms of the curvature and the amplitude of $J_{tt}$, making the existence of supercurrent vortices possible. Since varying $\alpha$ changes the topography of $J_{tt}$, it is reasonable that the vortex locations also changes. This can be seen in Figs.~\ref{fig:hxy3} and~\ref{fig:hxy4}, which show the current density distribution for $\alpha L = 1.2$ and $\alpha L = 1.6$, respectively. It is noted that for large enough Rashba parameter $\alpha$, no 0-$\pi$ transition will take place~\cite{arjoranta_prb_16}, and vortices will only appear for large exchange fields where the conventional contribution to the current, $J_I$, is sufficiently suppressed relative to $J_{tt}$.

Finally, we investigate the effect of varying the phase difference $\phi$ between the superconducting leads. In  Figs. \ref{fig:phi1} and \ref{fig:phi2} is shown the current density distribution for select values of $\phi$ for an exchange field of strength $h/\Delta = 1.0$ applied at an angle of $\theta = 45^{\circ}$ and a spin-orbit coupling strength of $\alpha L = 1.0$. It can be seen that the phase difference provides yet another means of enacting control over the vortices, with both their presence and location influenced. The two vortices translate vertically in opposite directions as the phase difference is increased. It is noted that the total current has a typical sinusoidal behavior, and that the current is zero for $\phi = 0$. 

Due to the complexity of the numerical problem, we have here focused on a particular system with a specific set of parameters which we believe to be experimentally relevant. However, the results can be generalized based on the current findings. For larger exchange fields (but not so large as to destroy the superconducting proximity effect) one can expect that the presence of vortices becomes more common, as the total current is gradually suppressed. Increasing the distance between the superconductors is also known to generate 0-$\pi$ transitions, around which one may expect vortices. The presence and location of the vortices may also be influenced by the width of the system, as is the case for flux induced vortices \cite{amundsen_scirep_16}. In other words, the results presented herein are not specific to the model considered. 

In an experimental setup, a slight misalignment in the orientation of an in-plane field will introduce an out-of-plane component which in turn generates an orbital effect that cannot in general be neglected. To investigate the degree to which this effect influences the current distribution, we have solved the system with an applied external flux $\ve{A}$. This produces an additional current contribution, given as
\begin{equation}
\ve{J}_A = 16\frac{J_0 \ve{A}}{\hbar^2}\int d\varepsilon\; \text{Im}(f_s\tilde{f}_s - \ve{f}_t\cdot\ve{\tilde{f}}_t)\tanh\frac{\beta\varepsilon}{2}
\end{equation}

\begin{figure}
	\subfloat[$\Phi/\Phi_0 = 0$]{%
		\includegraphics[width=0.5\columnwidth]{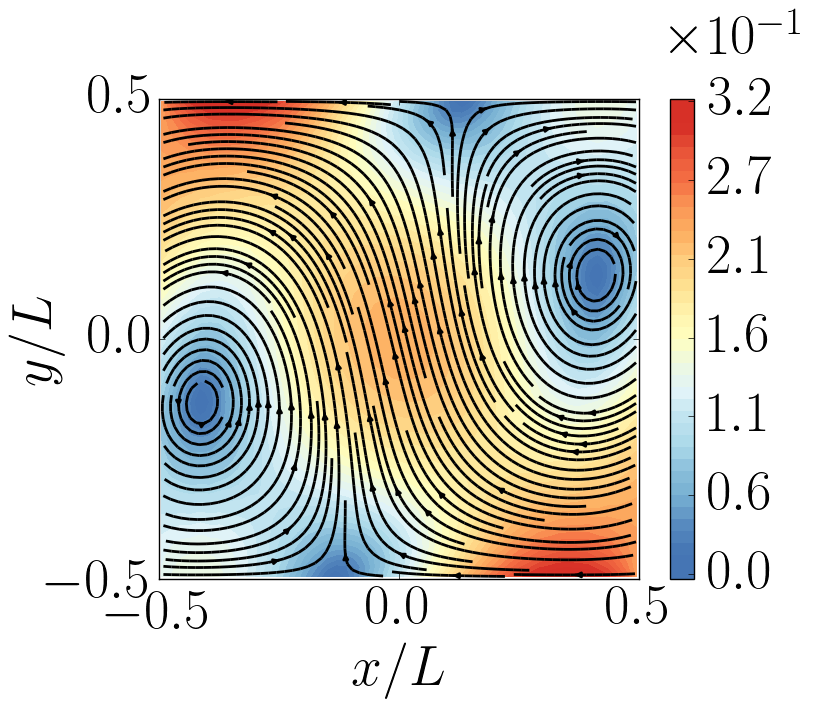}
		\label{fig:A1}
	}	
	\subfloat[$\Phi/\Phi_0 = 0.4$]{%
		\includegraphics[width=0.5\columnwidth]{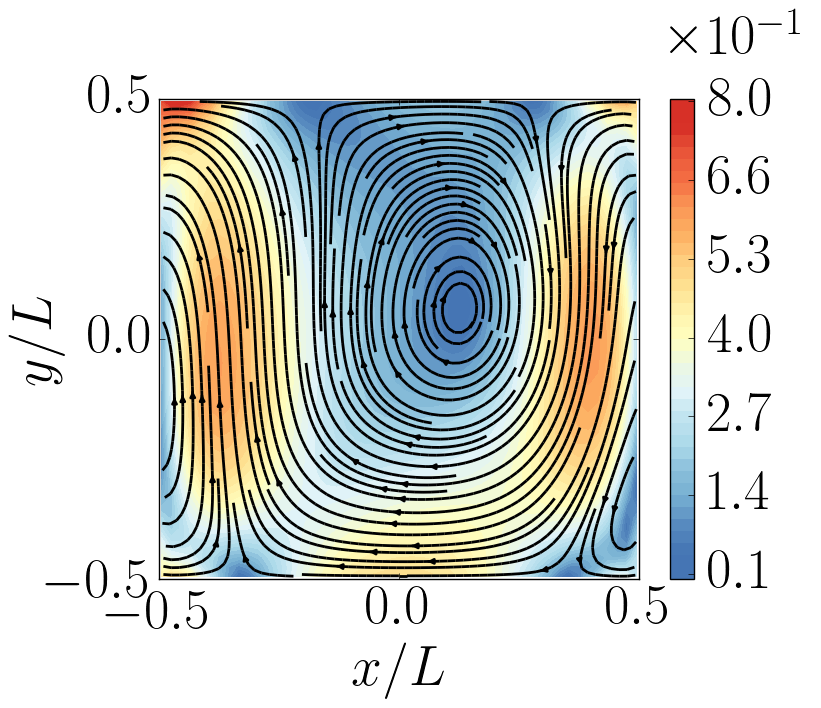}
		\label{fig:A2}
	}
	
	\subfloat[$\Phi/\Phi_0 = 1.2$]{%
		\includegraphics[width=0.5\columnwidth]{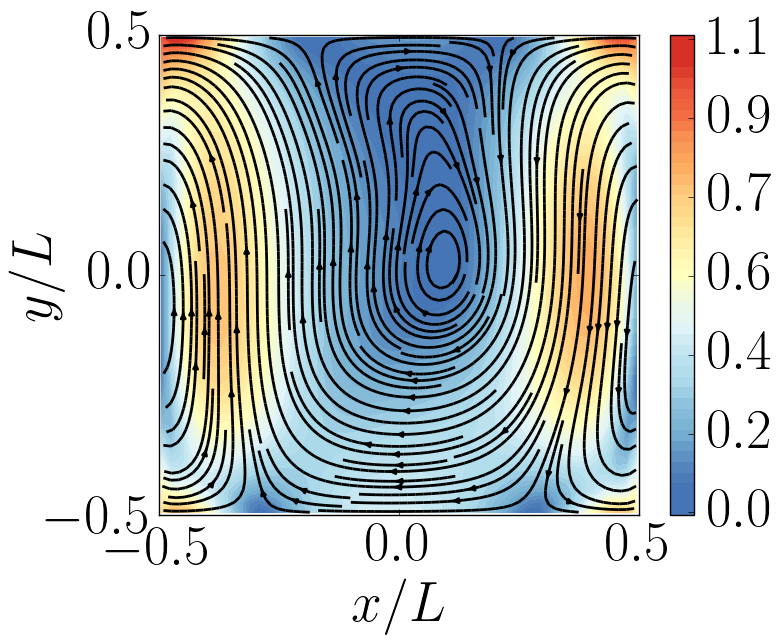}
		\label{fig:A3}
	}	
	\subfloat[$\Phi/\Phi_0 = 1.6$]{%
		\includegraphics[width=0.5\columnwidth]{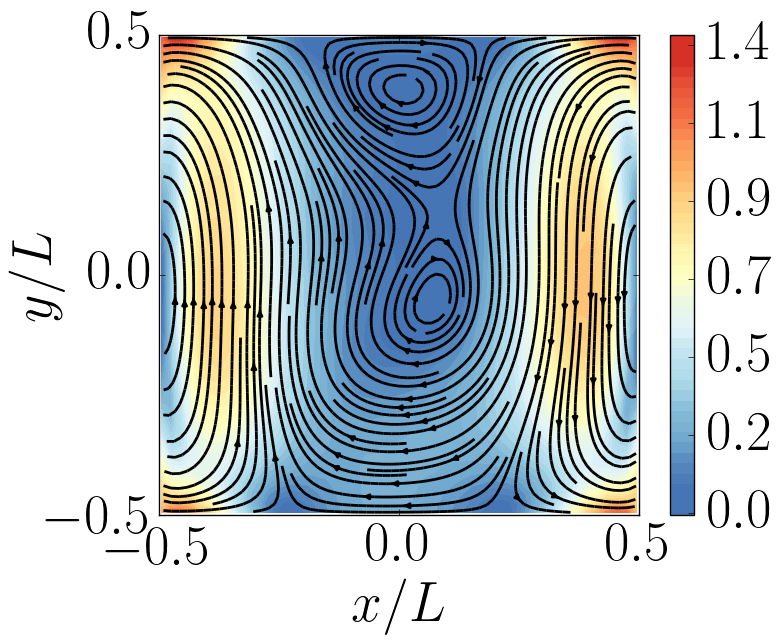}
		\label{fig:A4}
	}

	\caption{Streamlines of the charge current density $\ve{J}$ for increasing external flux $\Phi$ passing through the system in the $z$-direction. The exchange field strength is $h/\Delta = 1.0$, applied in an in-plane direction of $\theta = 45^{\circ}$, and the spin-orbit coupling strength is $\alpha L = 1.0$.}
	\label{fig:Acurrents}
\end{figure}

We use the Coulomb gauge, and define the vector potential as $\ve{A} = -B_{\perp}y\boldsymbol{e_x}$, where $\boldsymbol{e_x}$ is a unit vector pointing in the $x$-direction and $B_{\perp}$ is the out-of-plane component of the magnetic field. The flux passing through the system is then given by
$\Phi = B_{\perp}L^2$. In Figure \ref{fig:Acurrents} we show the current density distribution for various flux levels, with 
$\theta = 45^{\circ}$ and $h/\Delta = 1.0$. The strength of the spin-orbit coupling is $\alpha L = 1.0$. For $\Phi < \Phi_0$, where $\Phi_0 = \frac{h}{2e}$ is the flux quantum, there are no flux induced vortices, and we see from Figs. \ref{fig:Acurrents}a and \ref{fig:Acurrents}b that the effect of the external flux is to  translate the spin-orbit vortices towards the left, with the left-most vortex disappearing from the system. The right-most vortex eventually becomes trapped in the center of the junction for $\Phi \simeq \Phi_0$. For $\Phi > 1.2\Phi_0$, the vortex splits in two and aligns along the $y$-axis, as shown in Figs. \ref{fig:Acurrents}c and \ref{fig:Acurrents}d, which is typical for flux-induced vortices \cite{cuevas_prl_07}. At this point the spin-orbit induced vortices are indistinguishable from flux induced vortices.

To investigate whether spin-orbit vortices will be completely obscured by a small deviation from the in-plane orientation of the magnetic field in an experimental setting, we estimate an effective magnetic field from the exchange field via the Zeeman effect; $\ve{B}_\text{eff} = \frac{2\hbar}{\mu_B g}\ve{h}$, where $\mu_B$ is the Bohr magneton. We assume that the superconductors are niobium, for which the superconducting energy gap is given as ${\Delta \simeq \SI{1.5}{\milli\electronvolt}}$, and the diffusive coherence length is ${\xi\simeq \SI{15}{\nano\meter}}$. For a normal metal, where the $g$-factor is given as $g\simeq 2$, an exchange field strength of $|\ve{h}| = \Delta$ is produced by a magnetic field of $B_\text{eff} \simeq \SI{26}{\tesla}$. This magnitude is not intended as an experimentally feasible field, but is used here to show that even for huge external fields, the supercurrent vortex pattern predicted here remains robust toward an accidental out-of-plane field component. In some doped semiconductors, which are more relevant as candidate materials for the spin-orbit coupled region in the present study, the $g$-factor can be significantly higher, bringing the required field down to a more tractable level of order $\sim$ 1-\SI{2}{\tesla} \cite{hota_jpcm_91}. To estimate the out-of-plane component we use an angle equal to a realistic orientational uncertainty \cite{gusev_sst_99} of $\psi = 1^{\circ}$  , so that $B_{\perp} = B_\text{eff}\sin\psi\simeq$ \SI{0.45}{\tesla}. This amounts to a flux of $\Phi\simeq0.2\Phi_0$, which means that while the out-of-plane component will change the current distribution, it is small enough for the spin-orbit induced vortices to remain visible.

\section{Conclusion}
We have investigated two dimensional Josephson junctions with spin-orbit coupling, and find that vortices appear in the current density. The presence and location of these vortices may be tuned by varying either the exchange field strength, its direction, the strength of the spin-orbit coupling, or the phase difference. This "supercurrent vortex pinball effect" has its origin in the spin polarization of the triplet Cooper pairs induced in the system, and thus arises from what may be interpreted as a Cooper pair-induced inverse Edelstein effect.

\section*{Acknowledgements}
Funding via the “Outstanding Academic Fellows” program
at NTNU, the NV-Faculty, and
the Research Council of Norway Grant numbers 216700 and
240806, is gratefully acknowledged. We also acknowledge support from
NTNU and the Research Council of Norway for funding via
the Center of Excellence \textit{QuSpin.}

\end{document}